\documentstyle[aaspp4]{article}

\begin{document}

%

\title{Age, Metallicity, and the Distance to the Magellanic Clouds
From Red Clump Stars}

\author{Andrew A. Cole}
\affil{Department of Astronomy, University of Wisconsin-Madison \\
{\it cole@astro.wisc.edu}, 5534 Sterling Hall, 475 N. Charter
St., Madison, WI, 53706.}

\vspace{1.0in}

\begin{center}
Accepted for publication in {\it The Astrophysical Journal Letters}
\end{center}

\vspace{1.0in}

\begin{abstract}
We show that the luminosity dependence of the red clump stars
on age and metallicity can cause a difference of up to $\lesssim$0.6 mag
in the mean absolute I magnitude of the red clump between different stellar 
populations. We show that this effect may resolve the apparent
$\approx$0.4 mag discrepancy between red clump-derived distance
moduli to the Magellanic Clouds and those from, e.g., Cepheid variables.
Taking into account the population effects on red clump luminosity,
we determine a distance modulus to the LMC of 18.36 $\pm$ 0.17 mag,
and to the SMC of 18.82 $\pm$ 0.20 mag.  Our alternate red clump LMC
distance is consistent with the value (m-M)$_{LMC}$ = 18.50 $\pm$ 0.10 
adopted by the HST Cepheid Key Project.  We briefly examine model
predictions of red clump luminosity, and find that variations in 
helium abundance and core mass could bring the Clouds closer by some
0.10--0.15 mag, but not by the $\approx$0.4 mag that would result
from setting the mean absolute I-magnitude of the Cloud red clumps
equal to the that of the Solar neighborhood red clump.
\end{abstract}

\keywords{galaxies: distance and redshifts --- galaxies: individual 
(LMC, SMC) --- stars: late-type --- stars: evolution}

\newpage

\section{Introduction}

The distance to the Magellanic Clouds is a problem of great
astrophysical interest due to their role in the determination 
of extragalactic distances (c.f. \cite{mad98}). Despite the 
large amount of effort that has gone into the determination
of these distances, they remain a matter of some controversy
(see \cite{wes97}, \cite{wes90} for thorough discussions).

The success of the HST Cepheid Key Project, whose goal is the determination
of H$_0$ to an accuracy of 10\%, depends critically on knowledge
of the distance to the Clouds, especially the LMC.  Based on the
Cepheid variable Period-Luminosity relation (\cite{mad98}) and the
light echoes of SN1987A (\cite{pan97}), the Key Project has adopted
a distance modulus of (m-M)$_{LMC}$ = 18.50 $\pm$ 0.10 (50 $\pm$ 2 kpc)
(\cite{raw97}).
These determinations are in good agreement with recent derivations from
RR Lyrae variables, based on both ground-based, statistical parallax methods
(\cite{fea97b}), and on the {\sl Hipparcos}-calibrated distance scale
(\cite{gra97}, \cite{rei97}).

However, the LMC distance controversy is far from settled.  A substantial
fraction of recent techniques have yielded smaller values, typically
in the range of (m-M)$_{LMC}$ $\approx$ 18.30 $\pm$ 0.10.  These include
further analysis of the SN1987A light echoes (\cite{gou98}) and recalibration
of the RR Lyrae magnitude-metallicity relation (\cite{lay96}).

If the distance to the LMC is uncertain, that of the SMC is even more
so (\cite{wes97}, \cite{uda98}).  Because of its smaller Cepheid population
and its large line-of-sight depth, distance determinations to the SMC
are in generally poorly constrained.  We can only say with confidence
that it lies some 0.3--0.6 mag beyond the LMC.  The best Cepheid distance
to the SMC is 18.94 $\pm$ 0.04 (\cite{lan94}).

Recently, the large photometric sample of the OGLE microlensing survey
and its resulting high-quality color-magnitude diagrams of the Clouds
have permitted the development of the red clump of intermediate-age
helium-burning stars as a ``standard candle'' for single-step distance
determinations (\cite{pac98}).    
The key to this method has been the availability of accurate {\sl
Hipparcos} parallaxes to calibrate the absolute magnitude
of solar neighborhood red clump stars.  The {\sl Hipparcos} color-magnitude
diagrams (e.g., \cite{jim98}) show a solar-neighborhood red clump that 
has a very small dispersion in mean absolute I-band magnitude.  Because
the red clump is the dominant post-main-sequence evolutionary phase
for most stars, it makes a tempting target for the application of 
``standard candle'' techniques for distance determinations. 

The red clump method was developed very thoroughly and applied to
fields in the Galactic bulge (\cite{pac98}) and M31 (\cite{sta98}).
These studies used the mean absolute magnitude of solar-neighborhood
red clump stars, M$_I^0$ = $-$0.23 $\pm$ 0.03, obtained from a volume-limited
sample of 228 red clump stars observed with {\sl Hipparcos}.  
The red clump method is based on a well-populated, well-calibrated
phase of stellar evolution; in fact, it is possibly a more reliable
distance indicator than many other methods that have been employed.

In a recent paper, \cite{uda98} extended the red clump
method to the Magellanic Clouds.  After taking careful account of the
reddening distributions along Cloud lines of sight, they find mean red
clump magnitudes of I$_0$(LMC) = 17.85 $\pm$ 0.03, and
I$_0$(SMC) = 18.33 $\pm$ 0.03.  Using the solar neighborhood 
value M$_I^0$ = $-$0.23 $\pm$ 0.03, \cite{uda98} find distance moduli
of (m-M)$_{LMC}$ = 18.08 $\pm$ 0.15 and (m-M)$_{SMC}$ = 18.56 $\pm$ 0.09.
These values are $\approx$0.4 mag below the ``long'' distance scale
preferred by the HST Cepheid Key Project, and only marginally consistent
with the ``short'' scale.  \cite{sta98b} applied the same technique
to an independent large photometric survey of the LMC and obtained
the virtually identical result (m-M)$_{LMC}$ = 18.07 $\pm$ 0.12. 

\section{Nonstandard Candles}

As \cite{uda98} and \cite{sta98b} note, the question becomes: why
does such a robust method give results that are plainly inconsistent
with other well-developed techniques?  We suspect that the effects
of stellar evolution are responsible (see, e.g., \cite{cap95}, 
\cite{apa96}, \cite{jim98}, \cite{dac98}).

\cite{uda98} very briefly discuss the possibility that population 
differences between the solar neighborhood and the Clouds may be
responsible for the $\approx$0.4 mag distance discepancy they find.
However, they deem these stellar population effects to be negligible
for three main reasons (\cite{pac98}, \cite{sta98}). First, the
small observed dispersion in mean clump magnitude for each of the
{\sl Hipparcos}, Galactic bulge, M31, and Magellanic Cloud fields;
second, the lack of a trend in I$_0$ with (V$-$I) color in 
observed color-magnitude diagrams; and third, the small variation
in I$_0$ between fields with known metallicity differences in M31.

Each of these reasons is open to question.   If, for
example, the stellar populations differ {\it from each other} in their 
ages and metallicities, yet are homogeneous {\it within each field},
then each red clump may be intrinsically narrow, yet differ significantly
the solar neighborhood red clump.  We found an extreme example of
this effect in the very metal-poor, young dwarf irregular galaxy
Leo A (\cite{tol98}), whose red clump must be $\approx$0.4 mag
more luminous than the {\sl Hipparcos} red clump in order to 
yield a distance that can be reconciled with the color-magnitude
diagram of that galaxy.  We will address the empirical constancy
of M$_I^0$ and the lack of variation in the M31 fields after 
considering the theoretical model predictions for red clump properties.

To explore the variation of the mean red clump magnitude M$_I^0$ with
age and metallicity, we examined the theoretical red clump models
of \cite{sei87a} (SDW).  These models give red clump star evolutionary
tracks for masses from 0.74--1.70 M$_{\sun}$ and metallicities 
Z = 0.001, 0.004, 0.01, and 0.02.  These models thus
span the ranges of red clump formation epochs (1--10 Gyr), 
and of Galactic, LMC, and SMC metallicities.  In their analysis of SMC
star clusters, \cite{dac98} note that newer models (e.g., \cite{jim98})
yield essentially indistinguishable results.

\cite{sei87b} used these models, compared with star clusters in the
Clouds, to constrain the Cloud distances.  Interestingly, they also
preferred a short distance to the Clouds.  However, they do not 
exclude a long distance modulus because of the problems which, then
as now, plague attempts to derive absolute red clump properties from
theoretical models.  The most vexing of these problems are the issue
of mass-loss at the tip of the red giant branch, and the effects of
convective overshooting from the cores of these stars (c.f. the 
excellent discussion of mass loss in \cite{jim98}).

In light of these difficulties, we take an empirical approach, in 
which the {\sl Hipparcos} calibration of M$_I^0$ is used as our 
starting point, and deviations from M$_I^0$(local) are calculated using the
relative shifts between the SDW models.  We define the
mean luminosity of the red clump, L$_0$, for each track by finding the 
points at which the model was evolving most slowly, i.e., was most
likely to be observed at.  For each mass, we made a linear fit to the
variation of L$_0$ with Z and examined the residuals.  Where the
zeropoint offset was larger than the rms residual, we broke the fit
into two pieces, one for the high-metallicity end and one for the low.
We found the best single line fit to all masses is $\delta$M$_I^0$ = 
(0.21 $\pm$ 0.07) [Fe/H].

We identified the {\sl Hipparcos}-calibrated red clump M$_I^0$ with 
the 1.2 M$_{\sun}$, Z = 0.02 model clump.  This is appropriate for
a solar-metallicity population with roughly constant star formation
rate over the past 10 Gyr.  We can then calculate the predicted 
M$_I^0$ for arbitrary stellar population mixes given the mean ages
and metallicities of those populations using the formula 
$\delta$M$_I^0$ = -2.5 $\delta$L$_0$. Because the effective temperatures
under consideration do not change by much, we ignore the effects of a
changing bolometric correction for now.  Note that this would be
grossly incorrect for populations of still lower metallicity, such as
Leo A's, in which the clump is offset to the blue.  First we
correct each population for its metallicity by adding the 
$\delta$M$_I^{0, [Fe/H]}$ determined from our linear fits.  Then we add
the appropriate $\delta$M$_I^{0, age}$ found from the luminosity differences
between the various masses in the SDW grid of solar-metallicity models.
Table 1 gives our adopted values of $\delta$M$_I^0$ for the masses and 
metallicities we have considered.  For stars less massive than $\approx$1.2
M$_{\sun}$, (i.e., populations older than $\approx$4--5 Gyr), the red
clump magnitude is nearly independent of stellar mass, but can still vary
significantly with metal abundance.

To apply these corrections to the {\sl Hipparcos} M$_I^0$, it is necessary
to assume a star formation history (SFH) for the target galaxy.  For the LMC,
we adopt the SFH of \cite{hol97}, in
which 73\% of the red clump is produced
between 2 and 10 Gyr ago with a mean metallicity Z = 0.001, and the remainder
derives from a 1--2 Gyr-old population with Z = 0.008.  For this star
formation history, we derive $\delta$M$_I^0$ = $-$0.32 $\pm$ 0.03, yielding
M$_I^0$(LMC) = $-$0.55 $\pm$ 0.04.  The color shift between a
solar-metallicity clump and the LMC clump prompts us to revise M$_I^0$(LMC)
downward by $\approx$ +0.04 mag due to the increasing I-band bolometric
correction (BC$_I$) for these slightly hotter stars,
giving M$_I^0$(LMC) = $-$0.51
$\pm$ 0.04.  To test the sensitivity of this value
to adopted SFH, we also examine a much more ``burstlike'' SFH (\cite{val96}),
in which 44\% of the red clump is made between 2 and 10 Gyr in the past
with Z = 0.004, and 56\% of the clump is 1--2 Gyr old at Z = 0.008.  In
this scenario, M$_I^0$(LMC) = $-$0.46 $\pm$ 0.04.  The brighter red clump
produced by a younger population is somewhat offset by the fainter red
clump resulting from the higher metallicity in the Vallanari {\it et al.}
scenario.  In light of recent HST observations (\cite{geh98}), we consider
the LMC to be more ``Holtzman-like'' than ``Vallenari-like'', and adopt
M$_I^0$(LMC) = $-$0.51 $\pm$ 0.04.  This interpretation is reinforced
by the relatively small observed spread in red clump magnitude observed
by \cite{sta98b}, who point out that a bursting SFH
would lead to a dispersion in clump properties.  \cite{sta98b} note 
the relatively blue color of the LMC red clump and conclude that its
stars are metal-poor but not exceptionally young, in agreement with the
\cite{hol97} and \cite{geh98} models.

For the SMC, the SFH is much less constrained.  We consider two scenarios,
one with a constant star formation rate from 1--10 Gyr ago, and one in 
which all clump stars were produced in a burst 1--3 Gyr in the past (see
\cite{pag98}).  Adopting the age-metallicity relation observed in SMC star
clusters (\cite{dac98}), the mean SMC metallicities we use are, 
respectively, [Fe/H] = $-$1 and [Fe/H] = $-$0.8.  The constant SFR 
scenario yields $\delta$M$_I^0$ = $-$0.26 $\pm$ 0.04 and M$_I^0$(SMC) = 
$-$0.49 $\pm$ 0.06.  For the burst scenario we obtain
$\delta$M$_I^0$ = $-$0.36
$\pm$ 0.04 and M$_I^0$(SMC) = $-$0.59 $\pm$ 0.06.  These low-metallicity
models require a change in BC$_I$ of $\approx$ +0.05 from
the solar-metallicity models.  At the present time, either SFH alternative
seems plausible, and so we adopt a mean value of M$_I^0$(SMC) = $-$0.49
$\pm$ 0.06 (random) $\pm$ 0.05 (systematic).

Combining these new values of M$_I^0$ with the values of I$_0$
from \cite{uda98}, we find longer distances to the Magellanic Clouds:
(m-M)$_{LMC}$ = 18.36 $\pm$ 
0.05 (random) $\pm$ 0.12 (systematic); (m-M)$_{SMC}$ = 18.82 $\pm$ 0.07
(random) $\pm$ 0.13 (systematic).

How can we reconcile the variability of red clump properties with
the apparent constancy of such stars as observed in the {\sl Hipparcos}
and OGLE color-magnitude diagrams?   Apart from the homogeneity 
arguments given above, we can exploit the properties of the theoretical
models and the behavior of real stellar populations for some 
plausible explanations.

It is known from models (e.g., \cite{sei87a}) that metallicity
influences the temperature as well as the luminosity of the red clump stars;
metal-poor stars are predicted to be bluer as well as brighter
than their metal-rich counterparts.  The observed lack of variation
in I$_0$ as a function of color has thus been interpreted as
empirical proof that I$_0$ does not vary appreciably with metallicity
(Udalski, private communication).  The differing behavior of bolometric
luminosity and I-magnitude is presumed to arise from the increasing
BC$_I$ which must be applied to bluer stars.  However, we find that
BC$_I$ changes by only $\approx$ +0.05 mag between a solar-metallicity
clump and an SMC-metallicity clump.  This counteracts some of the
age and metallicity variation of M$_I^0$, especially for stars near
1 M$_{\sun}$, but cannot negate it completely.

The implicit assumption in the standard candle argument is that 
a range in metallicity is the major cause of a spread 
in red clump color.  This is almost
certainly untrue in a real stellar population, in which star formation
has proceeded across a finite range in time.  A real red clump contains
stars of many masses; each of these begins its helium-burning lifetime at a
color determined in part by its individual mass-loss history, and evolves
in color during its $\sim$10$^7$ year lifetime.  There is no unique mapping
from the color-magnitude diagram to initial mass and metallicity.

Another effect that must be considered
is the increase in model luminosities with both helium abundance Y and
core mass M$_c$ (\cite{swe78}, \cite{sei87a}, \cite{jim98}).   We considered
models of constant Y and M$_c$.  However, in real
stars,  Y increases with Z, and M$_c$
decreases with increasing Y (\cite{swe78}).  Thus we
might expect the mean clump luminosity to depend less strongly on
metallicity than we have calculated here. 

The exact variations of
M$_c$ and Y with Z are subject to much uncertainty, but we estimate
using the extensive tabulations of in \cite{swe78} that the effect is
likely to be $\delta$M$_I^{0, (M_c, Y)}$ $\lesssim$ 0.15 mag in most
cases, and $\lesssim$ 0.10 mag for the cases considered here.
These pieces of stellar physics are of the correct magnitude and sign
to erase the trend of M$_I^0$ with [Fe/H] for populations older than
$\approx$ 6 Gyr, and reduce its effect for younger populations.  
Because the {\sl Hipparcos} red clump (\cite{jim98}), the Galactic
bulge clump (\cite{pac98}) and the M31 halo (\cite{sta98}) are all
older populations, their approximately constant M$_I^0$ is relatively
unsurprising.  The younger Magellanic Cloud populations, especially
for the ``burst'' SFH scenarios retain their brighter clumps even
assuming a high helium content and weak coupling between M$_c$ and Y,
although $\delta$M$_I^0$ is reduced somewhat.   

Following the relations from the calculations of \cite{swe78} and 
adopting dY/dZ = 2.5 (\cite{jim98}), we
estimate that the effects of variation in Y and M$_c$ with Z can
dim the LMC's red clump by 0.13 mag for the \cite{hol97} SFH.  The
age-metallicity tradeoff can be seen strongly in the SMC. For our
constant SFR model, its clump will be dimmed by 0.15 mag relative
to the values above; for the burst SFH, its clump dims by only 0.04
mag.  With these numbers, our revised distances become (m-M)$_{LMC}$
= 18.24 $\pm$ 0.17, and (m-M)$_{SMC}$ = 18.73 $\pm$ 0.23.  However,
we have taken the variations in Y and M$_c$ from disparate
sources and these values merely serve to illustrate the potential
countervailing effects of stellar physics on $\delta$M$_I^0$.  More detailed 
modelling is required.

\section{Conclusion}

Our alternate red clump distance modulus (m-M)$_{LMC}$ = 18.36 $\pm$ 0.17
is $\approx$0.3 mag longer than that obtained under the
assumption that the LMC red clump mimics the Galaxy's.  This is
consistent with the ``long'' distance modulus of 18.50 $\pm$ 0.10 
adopted by the Cepheid Key Project, and, contrary to \cite{uda98}
and \cite{sta98b},  
in agreement with the most recent calibrations of the Cepheid
period-luminosity relation, which give (m-M)$_{LMC}$ = 18.44 $\pm$ 0.35
or 18.57 $\pm$ 0.11 (\cite{mad98}).  Our value of 18.36 $\pm$ 0.17 is
also in agreement with the shorter distance from RR Lyrae stars of
(m-M)$_{LMC}$ = 18.28 $\pm$ 0.13 (\cite{lay96}).

Taking account of the stellar population of the SMC pushes that galaxy
slightly farther away as well, from (m-M)$_{SMC}$ = 18.56 $\pm$ 0.09
derived by \cite{uda98}, to (m-M)$_{SMC}$ = 18.82 $\pm$ 0.20.
The larger value is in good agreement with that from
the Cepheid period-luminosity relation, (m-M)$_{SMC}$ = 18.94 $\pm$ 0.04
(\cite{lan94}).

The detailed stellar physics of mass-loss, and the relations between
Z, Y, and M$_c$ introduce significant uncertainty into the determination
of red clump absolute magnitudes, even in the I band.  It is quite 
likely that these effects may work to bring the Magellanic Clouds to 
a distance more consistent with the ``short'' (RR Lyrae) distance
scale than the ``long'' (Cepheid) scale.   However,
the red clump method does not require Magellanic Cloud distances
as much as 15\% smaller than commonly accepted.  Figure 1 shows the
menagerie of recent LMC distance determinations with errorbars, and 
it can be seen that the red clump method gives results consistent
with most of the other determinations.

We conclude that the red clump is indeed an extremely useful distance
indicator, as described by (\cite{pac98}, \cite{sta98}, and \cite{uda98}).
However, like most stellar ``standard candles'', its properties vary with the
composition and age of the host galaxy, and the assumption that all
red clumps are identical to the {\sl Hipparcos} red clump is
probably incorrect.   Among populations dominated by stars older
than $\approx$6 Gyr, the standard candle approximation should be valid
to a high degree for M$_I^0$.  For younger populations, M$_I^0$ is 
probably brighter than M$_I^0$(local), and the correction may amount
to as much as several tenths of a magnitude.

\acknowledgments

I would like to thank Jay Gallagher for many helpful discussions on 
the extragalactic distance scale and the populations of dwarf galaxies,
Andrzej Udalski for his comments on an earlier draft of this paper
that significantly improved the final version, and Bob Bless for his
comments on bolometric corrections.  I would also like to thank Brad
Gibson for his insightful comments.  This work has been supported as 
part of the Wide Field Planetary Camera 2 Investigation Definition Team's
study of Magellanic Cloud stellar populations.  The WFPC2 IDT is funded by
NASA through JPL under contract NAS 7--1260.

{}

\clearpage 

\begin{deluxetable}{ccc}
\tablewidth{0pt}
\tablenum{1}
\tablecaption{The shifts in mean red clump I-band absolute magnitude
as a function of mass and metallicity for the models of Seidel,
Demarque \& Weinberg (1987).  These shifts must be applied to the
{\sl Hipparcos} absolute magnitude calibration of the solar neighborhood
red clump (M$_I^0$(local) = $-$0.23 $\pm$ 0.03) before using the red 
clump as a standard candle for distance determinations.   The values
here will be diminished slightly by the increase in BC$_I$ of 
$\lesssim$ 0.05 mag caused by the correlation between metallicity and
clump color.  These values are computed with constant M$_c$ and Y; see
the text for the possible effects of realistic variation of these
quantities.}
\tablehead{
\colhead{Mass} &
\colhead{Z} &
\colhead{$\delta$M$_I^0$ (mag)}} 
\startdata
1.70	& 0.02	& -0.17 \nl
\nodata	& 0.01	& -0.27	\nl
\nodata & 0.004	& -0.42	\nl
\nodata	& 0.001	& -0.65 \nl
\hline
1.50	& 0.02	& -0.13	\nl
\nodata	& 0.01	& -0.23	\nl
\nodata	& 0.004	& -0.35	\nl
\nodata	& 0.001	& -0.43	\nl
\hline
1.20	& 0.02	& -0.00	\nl
\nodata	& 0.01	& -0.10	\nl
\nodata	& 0.004	& -0.20	\nl
\nodata	& 0.001	& -0.32	\nl
\hline
1.00	& 0.02	& +0.05	\nl
\nodata	& 0.01	& -0.03	\nl
\nodata	& 0.004	& -0.07	\nl
\nodata	& 0.001	& -0.10	\nl
\enddata
\end{deluxetable}

\clearpage

\begin{figure}
\plotone{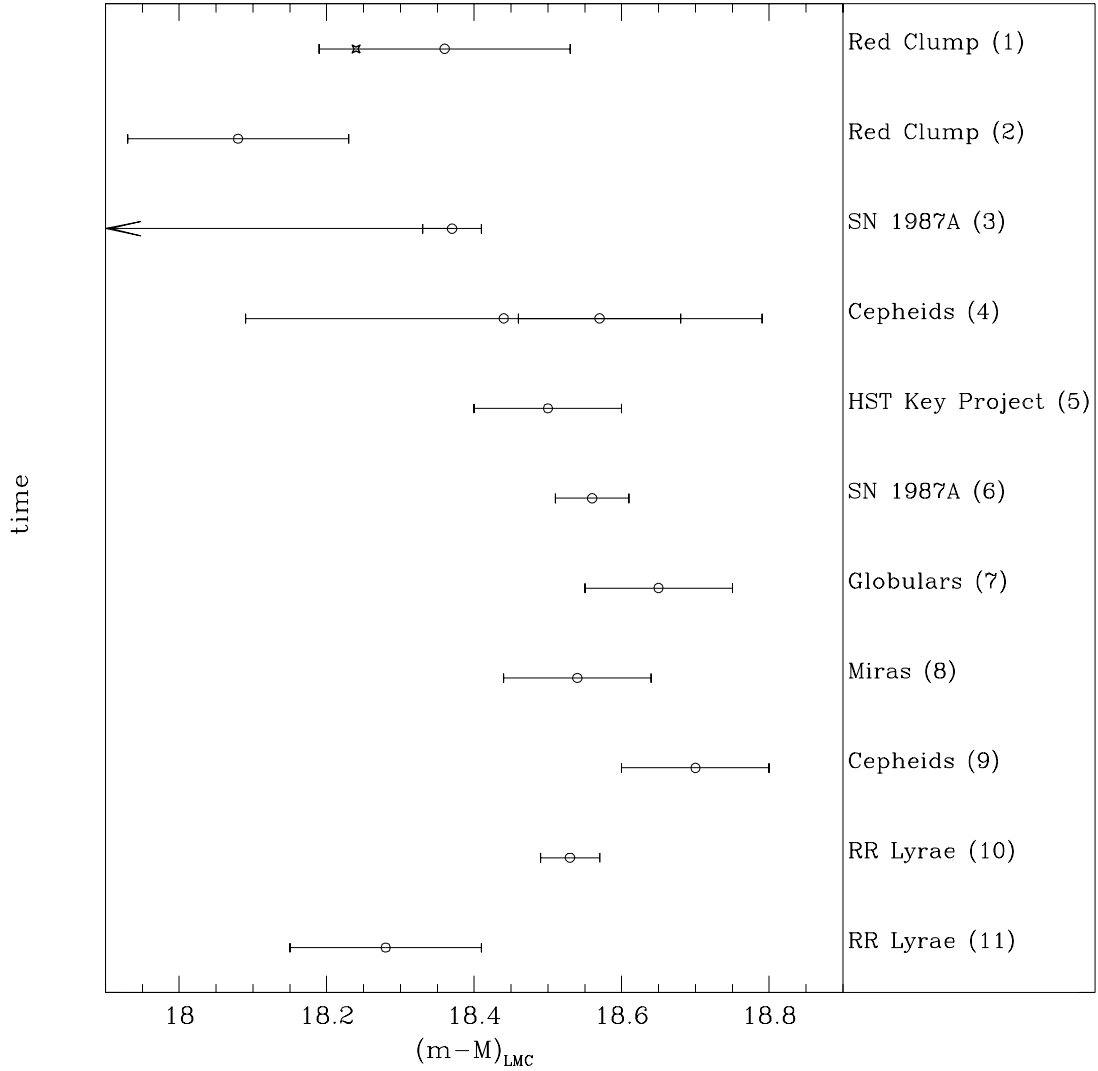}
\caption{Recent distance determinations to the LMC, ordered
chronologically from bottom to top.  The red clump method yields a range
of distances consistent with both Cepheid and RR Lyrae distance scales.
The {\bf x} on our Red Clump errorbar shows a potential shortening of the
distance scale based on the interplay of M$_c$ and Y in theoretical models.
References: (1) this paper; (2) \cite{uda98}, \cite{sta98b}; (3) \cite{gou98};
(4) \cite{mad98}; (5) \cite{raw97}; (6) \cite{pan97}; (7) \cite{rei97};
(8) \cite{van97}; (9) \cite{fea97}; (10) \cite{fea97b}; (11) \cite{lay96}.}
\end{figure}



\clearpage

\begin{thebibliography}{}

\bibitem[Aparicio {\it et al.} 1996]{apa96} Aparicio, A., Gallart, C.,
	Chiosi, C., \& Bertelli, G. 1996, \apjl, 469, L97.
\bibitem[Caputo {\it et al.} 1995]{cap95} Caputo, F., Castellani, V.,
	\& Degl'Innocenti, S. 1995, \aap, 304, 365.
\bibitem[Da Costa \& Hatzidimitriou 1998]{dac98} Da Costa, G.S., \&
	Hatzidimitriou, D. 1998, \aj, {\it in press, astro-ph/9802008}.
\bibitem[Feast (1997)]{fea97b} Feast, M.W. 1997, \mnras, 284, 761.
\bibitem[Feast \& Catchpole (1997)]{fea97} Feast, M.W., \& Catchpole, R.M.
	1997, \mnras, 286, L1.
\bibitem[Geha {\it et al.} 1998]{geh98} Geha, M.C., Holtzman, J.A., 
	Mould, J.R., Gallagher, J.S., Watson, A.M., Cole, A.A.,
	Grillmair, C.J., Stapelfeldt, K.R., \& WFPC2 IDT 1998, 
	\aj, 115, 1045.
\bibitem[Gould \& Uza 1998]{gou98} Gould, A., \& Uza, O. 1998, 
	\apjl, 494, L118.
\bibitem[Gratton {\it et al.} (1997)]{gra97} Gratton, R.G., Fusi Pecci, F.,
	Carretta, E., Clementini, G., Corsi, C.E., \& Lattanzi, M. 1997,
	\apj, 491, 749.
\bibitem[Holtzman {\it et al.} (1997)]{hol97} Holtzman, J.A., {\it et al.}
	1997, \aj, 113, 656.
\bibitem[Jimenez {\it et al.} 1998]{jim98} Jimenez, R., \& Flynn, C.
	\& Kotoneva, E. 1997,
	\mnras, {\it in press, astro-ph/9709056}.
\bibitem[Laney \& Stobie 1994]{lan94} Laney, C.D., \& Stobie, R.S.
	1994, \mnras, 266, 441.
\bibitem[Layden {\it et al.} 1996]{lay96} Layden, A.C., Hanson, R.B.,
	Hawley, S.L., Klemola, A.R., \& Hanley, C.J. 1996, \aj, 112, 2110. 
\bibitem[Madore \& Freedman 1998]{mad98} Madore, B.F., \& Freedman, W.L.
	1998, \apj, 492, 110.
\bibitem[Paczy\'{n}ski \& Stanek 1998]{pac98} Paczy\'{n}ski, B., \&
	Stanek, K.Z. 1998, \apjl, 494, L219.
\bibitem[Panagia {\it et al.} 1997]{pan97} Panagia, N., {\it et al.}
	1997, in {\it The Extragalactic Distance Scale}, ed. M. Livio
	\& M. Donahue, STSci Symp. 10, (Cambridge Univ. Pr, Cambridge). 
\bibitem[Pagel \& Tautvai\v{s}ien\.{e} 1998]{pag98} Pagel, B.E., \&
	Tautvai\v{s}ien\.{e}, G. 1998, \mnras, {\it in press,
	astro-ph/9801221}.
\bibitem[Rawson {\it et al.} 1997]{raw97} Rawson, D.M., {\it et al.}
	1997, \apj, 490, 517.
\bibitem[Reid (1997)]{rei97} Reid, I.N. 1997, \aj, 114, 161.
\bibitem[Seidel {\it et al.} (1987b)]{sei87b} Seidel, E., Da Costa, 
	G.S., \& Demarque, P. 1987, \apj, 313, 192.
\bibitem[Seidel {\it et al.} 1987a]{sei87a} Seidel, E., Demarque, P., 
	\& Weinberg, D. 1987, \apjs, 63, 917.
\bibitem[Stanek \& Garnavich 1998]{sta98} Stanek, K.Z., \& Garnavich,
	P.M. 1998, \apjl, {\it in press, astro-ph/9710327}.
\bibitem[Stanek {\it et al.} (1998b)]{sta98b} Stanek, K.Z., Zaritsky, D.,
	\& Harris, J. 1998, \apjl, {\it submitted, astro-ph/9803181}.
\bibitem[Sweigart \& Gross (1978)]{swe78} Sweigart, A.V., \& Gross,
	P.G. 1978, \apjs, 36, 405.
\bibitem[Tolstoy {\it et al.} 1998]{tol98} Tolstoy, E., Gallagher, J.S., 
	Cole, A.A., Hoessel, J.G., Saha, A., Skillman, E., Dohm-Palmer, 
	R., Mateo, M., \& Hurley-Keller, D. 1998, {\it in preparation}.
\bibitem[Udalski {\it et al.} (1998)]{uda98} Udalski, A., Szyma\'{n}ski, M.,
	Kubiak, M., Pietrzy\'{n}ski, G., Wo\'{z}niak, P., \& \.{Z}ebru\'{n},
	K. 1998, AcA, {\it submitted, astro-ph/9803035}.
\bibitem[Vallenari {\it et al.} 1996]{val96} Vallenari, A., Chiosi, C.,
	Bertelli, G., Aparicio, A., \& Ortolani, S. 1996, \aap, 309, 767.
\bibitem[van Leeuwen {\it et al.} (1997)]{van97} van Leeuwen, F., Feast, M.W.,
	Whitelock, P.A., \& Yudin, B. 1997, \mnras, 287, 955.
\bibitem[Westerlund 1990]{wes90} Westerlund, B.E. 1990, Astr. \& Astrophys.
	Rev., 2, 29.
\bibitem[Westerlund 1997]{wes97} Westerlund, B.E. 1997, {\it The Magellanic
Clouds}, (Cambridge Univ. Pr., Cambridge).

\end{thebibliography}
\end{document}